\documentclass[
a4paper, 
10pt,    
twoside, 
]{LTJournalArticle}

\usepackage{amsmath}
\usepackage{amsfonts}
\usepackage{amssymb}
\usepackage{amstext}
\usepackage{amsthm}

\newtheorem{proposition}{Proposition}

\newcounter{RomanNumber}

\usepackage{algorithm}
\usepackage{algorithmic}
\usepackage{color}
\usepackage{makecell}
\usepackage{multicol}

\title{The ECME Algorithm Using Factor Analysis for DOA Estimation in Nonuniform Noise}

\author{Mingyan Gong
}
\date{}

\renewcommand{\maketitlehookd}{
	\begin{abstract}
		\noindent Factor analysis (FA) plays a critical role in psychometrics, econometrics, and statistics. Recently, maximum likelihood FA (MLFA) has been applied to direction of arrival (DOA) estimation in unknown nonuniform noise and a variety of iterative approaches have been developed. In particular, the Factor Analysis for Anisotropic Noise (FAAN) method proposed by Stoica and Babu has excellent convergence properties. In this article, the Expectation/Conditional Maximization Either (ECME) algorithm, an extension of the expectation-maximization algorithm, is designed again for MLFA by introducing new complete data, which can thus use two explicit formulas to sequentially update the estimates of parameters at each iteration and have excellent convergence properties. Theoretical analysis shows that the ECME algorithm has almost the same computational complexity at each iteration as the FAAN method. However, numerical results show that the ECME algorithm yields faster stable convergence and the convergence to the global optimum is easier. Importantly, MLFA is not the best choice for the subspace based DOA estimation in unknown nonuniform noise.

        \par\textbf{Keywords: } Array signal processing, direction of arrival (DOA) estimation, expectation-maximization (EM) algorithm, factor analysis, maximum likelihood.
	\end{abstract}
}

\begin{document}

\maketitle 

\section{Introduction}

Factor analysis (FA) has been a tool in psychology for over one hundred years \cite{spearman} and its purpose is to extract or estimate the low-rank covariance matrix from a measurement covariance matrix. To this end, several techniques have been developed, but the only one available, which provides an efficient and consistent estimator, is maximum likelihood FA (MLFA) \cite{lawley,basilevsky,mulaik}. Unfortunately, this ML based estimator cannot be expressed in closed form, which leads to a variety of iterative approaches \cite{jöreskog}. In particular, one of the most promising is the Expectation-Maximization (EM) algorithm \cite{dempster, rubin}.

The EM algorithm regards observations or measurements as the incomplete data and introduces the complete data containing missing (latent or unobserved) data. The complete data is so manipulable that its log-likelihood function (LLF) is more tractable than the incomplete-data or actual LLF. Accordingly, the EM algorithm only uses the complete-data LLF to efficiently update the estimates of parameters at each iteration, which can still increase the actual likelihood. To be specific, each iteration of the EM algorithm consists of an expectation step (E-step) and a maximization step (M-step). However, the M-step requires maximizing the expected complete-data LLF provided by the E-step, which causes two shortcomings: 1) slow convergence, and 2) difficult maximization in numerous complex cases. To make up for both shortcomings, the Expectation/Conditional Maximization Either (ECME) algorithm, an extension of the EM algorithm, has been proposed \cite{liu} and later applied to MLFA \cite{liu2}. However, a conditional M-step (CM-step) at its each iteration utilizes Newton-Raphson, instead of an explicit formula, to update the estimates of partial parameters by the actual LLF and thus is complex as well as unstable.

In addition to psychometrics, econometrics, statistics and so on, FA recently finds applications in array signal processing, especially including direction of arrival (DOA) estimation in unknown nonuniform noise \cite{liao,stoica}. The nonuniform noise assumption can let the signal received at an array be modeled by FA so that MLFA is employed to estimate the signal subspace for the subspace based DOA estimation \cite{krim}. For this purpose, the iterative approach called Iterative ML Subspace Estimation (IMLSE) is proposed in \cite{liao}. This method sequentially updates the estimates of parameters at each iteration by two explicit formulas and is computationally attractive. However, Stoica and Babu have shown that the IMLSE method cannot guarantee its monotonicity and may fail to converge \cite{stoica}. Afterwards, they propose the novel iterative approach called Factor Analysis for Anisotropic Noise (FAAN), which has excellent convergence properties.

In this article, the ECME algorithm is designed again for MLFA by introducing new complete data, which can thus use two explicit formulas to sequentially update the estimates of parameters at each iteration and have excellent convergence properties. Theoretical analysis shows that the ECME algorithm has almost the same computational complexity at each iteration as the FAAN method. However, numerical results show that the ECME algorithm yields faster stable convergence and the convergence to the global optimum is easier. Importantly, MLFA is not the best choice for the subspace based DOA estimation in unknown nonuniform noise.

\emph{Notations}: $(\cdot)^T$ and $(\cdot)^H$ mean the transpose and conjugate transpose operators, respectively. $\mathrm{Det}(\boldsymbol{\Sigma})$, $\mathrm{Tr}\{\boldsymbol{\Sigma}\}$, $\boldsymbol{\Sigma}^{1/2}$, and $\boldsymbol{\Sigma}^{-1}$ stand for the determinant, trace, square root, and inverse of square matrix $\boldsymbol{\Sigma}$. $\Sigma_{i,j}$ is the element in the $i$th row and $j$th column of $\boldsymbol{\Sigma}$. $\boldsymbol{\Sigma}\succ\mathbf{0}_N$ and $\boldsymbol{\Sigma}\succeq\mathbf{0}_N$ denote that the $N\times N$ matrix $\boldsymbol{\Sigma}$ is positive definite and positive semi-definite, respectively. $\mathbf{I}_N$ is the identity matrix of order $N$ and $\jmath$ is the imaginary unit. $\mathcal{C}(\boldsymbol{\Sigma})$ denotes the column space of $\boldsymbol{\Sigma}$. $\theta^{(k)}$ is the iterative value of $\theta$ obtained at the $k$th iteration of an algorithm.

\section{Problem Description}

We consider a uniform linear array (ULA) consisting of $N$ isotropic sensors for analytical simplicity. $M~(M<N)$ narrow-band source signals are assumed to impinge upon the array from far field and the distance between arbitrary two adjacent sensors is the half-wavelength of the source signals. Thus, the composite signal received at this ULA can be expressed as
\begin{equation}   \label{1}    
\mathbf{y}(t)=\sum_{m=1}^M\mathbf{a}(\theta_m)s_m(t)+\mathbf{v}(t)=\mathbf{A}\mathbf{s}(t)+\mathbf{v}(t),
\end{equation}
where $\theta_m\in(0,\pi)~(\mathrm{radian})$ represents the DOA of the $m$th source signal, $\mathbf{a}(\theta_m)=[1~e^{-\jmath\pi\cos(\theta_m)}~\cdots~e^{-(N-1)\jmath\pi\cos(\theta_m)}]^T\in\mathbb{C}^N$ is the steering vector of the $m$th source signal, and $\mathbf{A}=[\mathbf{a}(\theta_1)~\cdots~\mathbf{a}(\theta_M)]$ is the array manifold matrix. For convenience, let $\theta_1<\cdots<\theta_M$. Additionally, $\mathbf{s}(t)=[s_1(t)~\cdots~s_M(t)]^T\in\mathbb{C}^M$ is the source signal vector, $\mathbf{v}(t)$ is the spatially nonuniform Gaussian white noise vector with mean zero and diagonal covariance matrix $\mathbf{Q}=\mathrm{diag}\{\sigma^2_1,\dots,\sigma^2_N\}\succ\mathbf{0}_N$, i.e., $\mathbf{v}(t)\sim\mathcal{CN}(\mathbf{0},\mathbf{Q})$.

When the stochastic source signal model is adopted, $\mathbf{s}(t)\sim\mathcal{CN}(\mathbf{0},\mathbf{P})$, where $\mathbf{P}\succeq\mathbf{0}_M$ is the source covariance matrix. In \eqref{1}, $\mathbf{s}(t)$ and $\mathbf{v}(t)$ are assumed to be mutually uncorrelated, which leads to $\mathbf{y}(t)\sim\mathcal{CN}(\mathbf{0},\mathbf{C})$ with $\mathbf{C}=\mathbf{A}\mathbf{P}\mathbf{A}^H+\mathbf{Q}\succ\mathbf{0}_N$ \cite{krim}. In practice, we first sample $\mathbf{y}(t)$ and obtain independent and identically distributed snapshots as follows:
\begin{equation}    \label{2}   
\mathbf{y}(t)=\mathbf{A}\mathbf{s}(t)+\mathbf{v}(t),t=1,\dots,L
\end{equation}
where $L$ is the number of snapshots. Then, we obtain the DOA estimates $\hat{\theta}_m$'s by processing the snapshots $\mathbf{y}(t)$'s.

From the processing perspective of ML direction finding \cite{kay}, we need to write the LLF of $\mathbf{Y}$ with $\mathbf{Y}=[\mathbf{y}(1)~\cdots~\mathbf{y}(L)]$ as
\begin{eqnarray}       %
l(\boldsymbol{\Phi})&=&\ln p(\mathbf{Y};\boldsymbol{\Phi})={\sum}_{t=1}^L\ln p\big(\mathbf{y}(t);\boldsymbol{\Phi}\big)\nonumber\\
&=&a-L\Big(\ln\mathrm{Det}(\mathbf{C})+\mathrm{Tr}\{\hat{\mathbf{R}}\mathbf{C}^{-1}\}\Big)
\end{eqnarray}
where $a$ is a constant, $\boldsymbol{\Phi}=(\boldsymbol{\theta},\mathbf{P},\mathbf{Q})$ with $\boldsymbol{\theta}=(\theta_1,\dots,\theta_M)$, $p(\cdot)$ is the probability density function (PDF), and $\hat{\mathbf{R}}=(1/L)\sum_{t=1}^L\mathbf{y}(t)\mathbf{y}^H(t)\succ\mathbf{0}_N$\footnote{In this article, we assume $\hat{\mathbf{R}}\succ\mathbf{0}_N$, which requires $L\ge N$.} is the snapshot covariance matrix. Importantly, we need to solve the equivalent ML direction finding problem \cite{jaffer, stoica2, pesavento}:
\begin{equation}   \label{3}    
\min_{\boldsymbol{\theta}\in(0,\pi)^M,\mathbf{P}\succeq\mathbf{0}_M,\mathbf{Q}\succ\mathbf{0}_N}\ln\mathrm{Det}(\mathbf{C})+\mathrm{Tr}\{\hat{\mathbf{R}}\mathbf{C}^{-1}\}.
\end{equation}

\section{Subspace Estimation Using MLFA}

The ML direction finding problem \eqref{3} is intractable since $\mathbf{C}=\mathbf{APA}^H+\mathbf{Q}$ has excessive constraints on free parameters (unknowns) to be estimated. To alleviate this issue, we should release as many constraints as possible and introduce as many free parameters as possible. Thus, we aim not to solve problem \eqref{3} in this article.

\subsection{MLFA Estimation Problem}

It is well-known that the subspace based DOA estimation only requires the signal subspace $\mathcal{C}(\mathbf{A})$ \cite{krim}. As a result, let $\mathbf{C}$ be the population covariance matrix in FA \cite{basilevsky} 
\begin{equation}   \label{4}    
\mathbf{C}=\mathbf{S}\mathbf{S}^H+\mathbf{Q},
\end{equation}
where all the $NM$ elements in $\mathbf{S}\in\mathbb{C}^{N\times M}$ are free parameters (factor loadings) and hence $\mathbf{SS}^H$ must be able to equal any $\mathbf{APA}^H$. In this article, we also use $\mathbf{S}$ to estimate $\mathcal{C}(\mathbf{A})$. Eq. \eqref{4} makes problem \eqref{3} become the classic MLFA estimation problem \cite{lawley,basilevsky,mulaik,jöreskog}:
\begin{equation}   \label{5}    
\min_{\mathbf{S},\mathbf{Q}\succ\mathbf{0}_N}\ln\mathrm{Det}(\mathbf{C})+\mathrm{Tr}\{\hat{\mathbf{R}}\mathbf{C}^{-1}\}.
\end{equation}

For solving problem \eqref{5}, the IMLSE method utilizes a fixed-point iteration and sequentially (not simultaneously) updates the estimates of $\mathbf{S}$ and $\mathbf{Q}$ at each iteration by two explicit formulas \cite{liao}. However, Stoica and Babu have shown that the IMLSE method cannot guarantee its monotonicity and may fail to converge \cite{stoica}. Hence, they propose the FAAN method with excellent convergence properties. We outline the FAAN method in the next subsection.

\subsection{FAAN Method}

When a direct minimization over all free parameters is very difficult, an iterative (coordinate descent) approach or alternating minimization can always be adopted \cite{bezdek}. As mentioned before, the FAAN method has been proposed for problem \eqref{5} and its each iteration consists of two steps. Specifically, since $\mathbf{C}=\mathbf{SS}^H+\mathbf{Q}$ makes updating the estimates of $\mathbf{S}$ and $\mathbf{Q}$ difficult, a new parametrization of $\mathbf{C}$ is designed as
\begin{equation}   \label{6}    
\mathbf{C}=\mathbf{Q}^{1/2}(\mathbf{U}\boldsymbol{\Lambda}\mathbf{U}^H+\mathbf{I}_N)\mathbf{Q}^{1/2},
\end{equation}
where $\boldsymbol{\Lambda}\succeq\mathbf{0}_M$, $\mathbf{U}\boldsymbol{\Lambda}\mathbf{U}^H$ is the eigen-decomposition of $\mathbf{Q}^{-1/2}\mathbf{SS}^H\mathbf{Q}^{-1/2}$, and $\mathbf{S}=\mathbf{Q}^{1/2}\mathbf{U}\boldsymbol{\Lambda}^{1/2}$ for convenience. Thus, problem \eqref{5} is changed to
\begin{eqnarray}   \label{7}    
\min_{\mathbf{U}^H\mathbf{U}=\mathbf{I}_M,\boldsymbol{\Lambda}\succeq\mathbf{0}_M,\mathbf{Q}\succ\mathbf{0}_N}\ln\mathrm{Det}(\mathbf{Q})+\ln\mathrm{Det}(\boldsymbol{\Lambda}+\mathbf{I}_M)\nonumber\\
+\mathrm{Tr}\big\{\mathbf{Q}^{-1/2}\hat{\mathbf{R}}\mathbf{Q}^{-1/2}(\mathbf{U}\boldsymbol{\Lambda}\mathbf{U}^H+\mathbf{I}_N)^{-1}\big\}.
\end{eqnarray}

According to \eqref{7},  given $\mathbf{Q}=\mathbf{Q}^{(k-1)}\succ\mathbf{0}_N$, the first step at the $k$th iteration of the FAAN method \emph{simultaneously} obtains $\mathbf{U}^{(k)}$ and $\boldsymbol{\Lambda}^{(k)}$ by
\begin{eqnarray}   \label{8}    
\min_{\mathbf{U}^H\mathbf{U}=\mathbf{I}_M,
\boldsymbol{\Lambda}\succeq\mathbf{0}_M}\ln\mathrm{Det}(\boldsymbol{\Lambda}+\mathbf{I}_M)+\nonumber\\
\mathrm{Tr}\big\{\tilde{\mathbf{R}}^{(k)}(\mathbf{U}\boldsymbol{\Lambda}\mathbf{U}^H+\mathbf{I}_N)^{-1}\big\},
\end{eqnarray}
where $\tilde{\mathbf{R}}^{(k)}=[\mathbf{Q}^{(k-1)}]^{-1/2}\hat{\mathbf{R}}[\mathbf{Q}^{(k-1)}]^{-1/2}\succ\mathbf{0}_N$. This leads to
\begin{subequations} \label{9}
\begin{eqnarray}   \label{9a}    
\mathbf{U}^{(k)}&=&\mathbf{B}^{(k)}\in\mathbb{C}^{N\times M},  \\
\Lambda_{m,m}^{(k)}&=&(\lambda_m^{(k)}-1)_+\ge0,m=1,\dots,M\label{9b}
\end{eqnarray}
\end{subequations}
where the $\lambda_m^{(k)}$'s are the eigenvalues of $\tilde{\mathbf{R}}^{(k)}$ and $\lambda_1^{(k)}\ge\cdots\ge\lambda_N^{(k)}>0$, the $m$th column of $\mathbf{B}^{(k)}$ is the eigenvector corresponding to $\lambda_m^{(k)}$. Moreover, the operator $(b)_+$ replaces the negative value of real number $b$ with zero and leaves the nonnegative value unchanged.

The second step at the $k$th iteration of the FAAN method sequentially (\textit{one by one}) obtains $\sigma_1^{(k)},\dots,\sigma_N^{(k)}$ by the $N$ problems:
\begin{eqnarray}   \label{10}    
\min_{\sigma_n>0}\ln\mathrm{Det}(\mathbf{Q})+
\mathrm{Tr}\{\mathbf{Q}^{-1/2}\hat{{\mathbf{R}}}\mathbf{Q}^{-1/2}\boldsymbol{\Gamma}^{(k)}\},
\end{eqnarray}
where $n=1,\dots,N$ and $\boldsymbol{\Gamma}^{(k)}=\big(\mathbf{U}^{(k)}\boldsymbol{\Lambda}^{(k)}[\mathbf{U}^{(k)}]^H+\mathbf{I}_N\big)^{-1}\succ\mathbf{0}_N$\footnote{We do not adopt $\boldsymbol{\Gamma}^{(k)}=\mathbf{I}_N+\mathbf{U}^{(k)}\big([\boldsymbol{\Lambda}^{(k)}]^{-1}-\mathbf{I}_M\big)[\mathbf{U}^{(k)}]^H$ since numerical results show its low accuracy.}. This leads to
\begin{eqnarray}   \label{11}    
\sigma_n=\frac{1}{2}\Big(b_n^{(k)}+\sqrt{[b_n^{(k)}]^2+4c_n^{(k)}}\Big)>0
\end{eqnarray}
where $n=1,\dots,N$, $\Re(b)$ denotes the real part of complex number $b$, $b_n^{(k)}=\sum_{i\ne n}\Re\{\hat{R}_{i,n}\Gamma_{i,n}^{(k)}\}/\sigma_i$ and $c_n^{(k)}=\hat{R}_{n,n}\Gamma_{n,n}^{(k)}>0$. The details of the FAAN method are in \textbf{Algorithm \ref{a1}}.

\begin{algorithm}
\caption{FAAN Method} \label{a1}
\begin{algorithmic}[1]
\STATE {Initialize $\mathbf{Q}^{(0)}\succ\mathbf{0}_N$, $K_1$, $K_2$, and $k=1$.}
        \WHILE{$k\le K_1$}
           \STATE {Obtain $\mathbf{U}^{(k)}$ and $\boldsymbol{\Lambda}^{(k)}$ using \eqref{9}.}
           \STATE {Obtain $\mathbf{Q}^{(k)}$ using \eqref{11} $K_2$ times.}
           \STATE {$k=k+1$.}
       \ENDWHILE
\STATE {Let $\hat{\mathbf{S}}=[\mathbf{Q}^{(K_1)}]^{1/2}\mathbf{U}^{(K_1)}[\boldsymbol{\Lambda}^{(K_1)}]^{1/2}$ and output $\hat{\mathbf{S}}$.}
\end{algorithmic}
\end{algorithm}

\section{ECME Algorithm}

The ECME algorithm has been designed for problem \eqref{5} in \cite{liu2} and yields rapider convergence than the EM algorithm designed in \cite{rubin}. However, a CM-step at its each iteration utilizes Newton-Raphson, instead of an explicit formula, to update the estimate of $\mathbf{Q}$ by the objective function or negative LLF $f(\mathbf{S},\mathbf{Q})=\ln\mathrm{Det}(\mathbf{C})+\mathrm{Tr}\{\hat{\mathbf{R}}\mathbf{C}^{-1}\}$ in problem \eqref{5}, leading to that this step is complex and the monotonicity cannot be guaranteed. In this section, we again design the ECME algorithm by introducing new complete data, which can thus use two explicit formulas to sequentially update the estimates of $\mathbf{S}$ and $\mathbf{Q}$ at each iteration and have excellent convergence properties.

\subsection{Algorithm Procedure}

For problem \eqref{5}, we do not adopt the classic complete data $(\mathbf{Y},\mathbf{X})$ with $\mathbf{X}=\big[\mathbf{x}(1)~\cdots~\mathbf{x}(L)\big]$ in \cite{rubin,liu2} and instead introduce the new complete data $(\mathbf{X},\mathbf{V})$ with $\mathbf{V}=\big[\mathbf{v}(1)~\cdots~\mathbf{v}(L)\big]$\footnote{The complete-data LLF of $(\mathbf{Y},\mathbf{X})$ is
\begin{eqnarray} &l(\mathbf{X};\mathbf{S},\mathbf{Q})=\ln p(\mathbf{Y},\mathbf{X};\mathbf{S},\mathbf{Q})\nonumber\\&={\sum}_{t=1}^L\Big[\ln p\big(\mathbf{y}(t)\mid\mathbf{x}(t);\mathbf{S},\mathbf{Q}\big)+\ln p\big(\mathbf{x}(t)\big)\Big]\nonumber\\
    &=c-L\ln\mathrm{Det}(\mathbf{Q})-{\sum}_{t=1}^L\big[\mathbf{y}(t)-\mathbf{Sx}(t)\big]^H\mathbf{Q}^{-1}\big[\mathbf{y}(t)-\mathbf{Sx}(t)\big]\nonumber
\end{eqnarray}
where $c$ is a term unrelated to any free parameters in $(\mathbf{S},\mathbf{Q})$, and more complex than that of $(\mathbf{X},\mathbf{V})$ in \eqref{13}. Numerical results show that the ECME algorithm with $(\mathbf{X},\mathbf{V})$ yields the same convergence as that with $(\mathbf{Y},\mathbf{X})$, but the ECME algorithm with $(\mathbf{X},\mathbf{V})$ needs fewer computations at each iteration.}. Toward this end, we rewrite the snapshots or measurements $\mathbf{Y}$ in \eqref{2} by \cite{lawley,basilevsky,mulaik}
\begin{eqnarray}   \label{12}    
\mathbf{y}(t)=\mathbf{S}\mathbf{x}(t)+\mathbf{v}(t),t=1,\dots,L
\end{eqnarray}
where $\mathbf{x}(t)\sim\mathcal{CN}(\mathbf{0},\mathbf{I}_M)$ is composed of $M$ common factors and uncorrelated with $\mathbf{v}(t)$. Here, the statistical design of $\mathbf{x}(t)$ is based on $\mathbf{y}(t)\sim\mathcal{CN}(\mathbf{0},\mathbf{SS}^H+\mathbf{Q})$. Then, the complete-data LLF is
\begin{eqnarray}  \label{13}     
l(\mathbf{X},\mathbf{V};\mathbf{Q})&=&\ln p(\mathbf{X},\mathbf{V};\mathbf{Q})=\ln p(\mathbf{X})+\ln p(\mathbf{V};\mathbf{Q})\nonumber\\
&=&c-L\Big(\ln\mathrm{Det}(\mathbf{Q})+\mathrm{Tr}\{\hat{\mathbf{R}}_v\mathbf{Q}^{-1}\}\Big)
\end{eqnarray}
where $c$ is a term unrelated to any free parameters in $(\mathbf{S},\mathbf{Q})$ and $\hat{\mathbf{R}}_{v}=(1/L)\sum_{t=1}^L\mathbf{v}(t)\mathbf{v}^H(t)$. With \eqref{13}, we can construct the ECME algorithm, whose E-step and two CM-steps at the $k$th iteration are derived below.

\subsubsection{First CM-step:} Given $\mathbf{Q}=\mathbf{Q}^{(k-1)}\succ\mathbf{0}_N$, let this step only obtain $\mathbf{S}^{(k)}$ by directly minimizing the objective function $f(\mathbf{S},\mathbf{Q})$ in problem \eqref{5} with respect to $\mathbf{S}$, i.e.,
\begin{equation}   \label{14}    
\min_{\mathbf{S}}\ln\mathrm{Det}\big(\mathbf{SS}^H+\mathbf{Q}^{(k-1)}\big)+\mathrm{Tr}\big\{\hat{\mathbf{R}}(\mathbf{SS}^H+\mathbf{Q}^{(k-1)})^{-1}\big\}.
\end{equation}
This problem is classic \cite{lawley,basilevsky,mulaik} but the classic solutions in \cite{joreskog,bartholomew} may not exist. Thus, we resort to the FAAN method by
\begin{eqnarray}   \label{15}
\mathbf{S}^{(k)}=[\mathbf{Q}^{(k-1)}]^{1/2}\mathbf{U}^{(k)}[\boldsymbol{\Lambda}^{(k)}]^{1/2}\in\mathbb{C}^{N\times M},
\end{eqnarray}
where $\mathbf{U}^{(k)}$ and $\boldsymbol{\Lambda}^{(k)}$ are from \eqref{9}.

\subsubsection{E-step:}Given $\mathbf{Y}$, $\mathbf{S}^{(k)}$, and $\mathbf{Q}^{(k-1)}$, this step finds the conditional expectation of the complete-data LLF $l(\mathbf{X},\mathbf{V};\mathbf{Q})$ in \eqref{13} by
\begin{eqnarray}  \label{16}     
&l(\mathbf{Q};\boldsymbol{\Omega}^{(k)})=\mathrm{E}\big\{l(\mathbf{X},\mathbf{V};\mathbf{Q})\mid\mathbf{Y};\boldsymbol{\Omega}^{(k)}\}\nonumber\\
&=c-L\Big(\ln\mathrm{Det}(\mathbf{Q})+\mathrm{Tr}\{\hat{\mathbf{R}}_v^{(k)}\mathbf{Q}^{-1}\}\Big)
\end{eqnarray}
where $\boldsymbol{\Omega}^{(k)}=(\mathbf{S}^{(k)},\mathbf{Q}^{(k-1)})$, $\mathrm{E}\{\cdot\}$ and $\mathrm{D}\{\cdot\}$ mean the expectation and covariance operators, respectively. Moreover,
\begin{eqnarray} \label{16e}
\hat{\mathbf{R}}^{(k)}_{v}&=&\mathrm{E}\big\{\hat{\mathbf{R}}_{v}\mid\mathbf{Y};\boldsymbol{\Omega}^{(k)}\big\}\nonumber\\
&=&\frac{1}{L}\sum_{t=1}^L\mathrm{E}\big\{\mathbf{v}(t)\mathbf{v}^H(t)\mid\mathbf{Y};\boldsymbol{\Omega}^{(k)}\big\}\nonumber\\
&=&\frac{1}{L}\sum_{t=1}^L\Big(\boldsymbol{\Sigma}^{(k)}+\boldsymbol{\mu}^{(k)}(t)\big[\boldsymbol{\mu}^{(k)}(t)\big]^H\Big)\nonumber\\
&=&\boldsymbol{\Sigma}^{(k)}+\frac{1}{L}\sum_{t=1}^L\boldsymbol{\mu}^{(k)}(t)[\boldsymbol{\mu}^{(k)}(t)]^H\nonumber\\
&=&\boldsymbol{\Delta}^{(k)}+[\boldsymbol{\delta}^{(k)}]^H\hat{\mathbf{R}}\boldsymbol{\delta}^{(k)}\succ\mathbf{0}_N
\end{eqnarray}
where $\mathbf{C}^{(k)}=\mathbf{S}^{(k)}[\mathbf{S}^{(k)}]^H+\mathbf{Q}^{(k-1)}\succ\mathbf{0}_N$, $\boldsymbol{\delta}^{(k)}=[\mathbf{C}^{(k)}]^{-1}\mathbf{Q}^{(k-1)}$\footnote{We do not adopt $[\mathbf{C}^{(k)}]^{-1}=\mathbf{q}^{(k)}-\mathbf{q}^{(k)}\mathbf{S}^{(k)}(\boldsymbol{\Lambda}^{(k)}+\mathbf{I}_M)^{-1}[\mathbf{S}^{(k)}]^H\mathbf{q}^{(k)}$ with $\mathbf{q}^{(k)}=[\mathbf{Q}^{(k-1)}]^{-1}$ since numerical results show its low accuracy.}, and $
\boldsymbol{\Delta}^{(k)}=\mathbf{Q}^{(k-1)}-\mathbf{Q}^{(k-1)}\boldsymbol{\delta}^{(k)}\succeq\mathbf{0}_N$. Additionally, the conditional distribution of $\mathbf{v}(t)$ can be derived from \cite{rhodes}, and
\begin{subequations} 
\begin{eqnarray}
\boldsymbol{\mu}^{(k)}(t)&=&\mathrm{E}\big\{\mathbf{v}(t)\mid\mathbf{Y};\boldsymbol{\Omega}^{(k)}\big\}
=[\boldsymbol{\delta}^{(k)}]^H\mathbf{y}(t),\\
\boldsymbol{\Sigma}^{(k)}&=&\mathrm{D}\big\{\mathbf{v}(t)\mid\mathbf{Y};\boldsymbol{\Omega}^{(k)}\big\}
=\boldsymbol{\Delta}^{(k)}\succeq\mathbf{0}_N.
\end{eqnarray}
\end{subequations} 

\subsubsection{Second CM-step:}This step only obtains $\mathbf{Q}^{(k)}$ by maximizing the expected complete-data LLF $l(\mathbf{Q};\boldsymbol{\Omega}^{(k)})$ in \eqref{16} with respect to $\mathbf{Q}$, i.e.,
\begin{eqnarray}  \label{17}     
\min_{\mathbf{Q}\succ\mathbf{0}_N}\ln\mathrm{Det}(\mathbf{Q})+\mathrm{Tr}\big\{\hat{\mathbf{R}}_v^{(k)}\mathbf{Q}^{-1}\}.
\end{eqnarray}
This problem can be decomposed into the $N$ \textit{parallel} subproblems:
\begin{eqnarray}  \label{18}     
\min_{\sigma_n^2>0}\ln(\sigma_n^2)+[\hat{\mathbf{R}}_v^{(k)}]_{n,n}/\sigma^2_n,n=1,\dots,N.
\end{eqnarray}
Let $f(\sigma^2_n)=\ln(\sigma^2_n)+[\hat{\mathbf{R}}_v^{(k)}]_{n,n}/\sigma^2_n$ with $[\hat{\mathbf{R}}_v^{(k)}]_{n,n}>0$ and its derivative is
\begin{eqnarray}
    f'(\sigma^2_n)&=&\frac{1}{\sigma^2_n}-\frac{[\hat{\mathbf{R}}_v^{(k)}]_{n,n}}{\sigma^4_n}=\frac{\sigma^2_n-[\hat{\mathbf{R}}_v^{(k)}]_{n,n}}{\sigma^4_n}\nonumber\\
&&\left\{
\begin{array}{ll}
<0, & 0<\sigma^2_n<[\hat{\mathbf{R}}_v^{(k)}]_{n,n}, \\
{=0,} & {\sigma^2_n=[\hat{\mathbf{R}}_v^{(k)}]_{n,n}}, \\
{>0,} & {\sigma^2_n>[\hat{\mathbf{R}}_v^{(k)}]_{n,n}}, \\
\end{array}
\right.
\end{eqnarray}
which indicates that when $\sigma^2_n=[\hat{\mathbf{R}}_v^{(k)}]_{n,n}$, $f(\sigma^2_n)$ reaches the minimum. Thus,
the solutions in subproblems \eqref{18} are
\begin{eqnarray}   \label{19}    
[\sigma^2_n]^{(k)}&=&[\hat{\mathbf{R}}_v^{(k)}]_{n,n}>0,n=1,\dots,N,\nonumber\\
\Rightarrow\mathbf{Q}^{(k)}&=&\mathrm{diag}\big(\hat{\mathbf{R}}_{v}^{(k)}\big)\succ\mathbf{0}_N,
\end{eqnarray}
where $\mathrm{diag}(\hat{\mathbf{R}}_{v}^{(k)})$ is the diagonal matrix composed of the diagonal part of $\hat{\mathbf{R}}_{v}^{(k)}$. The details of the ECME algorithm are in \textbf{Algorithm \ref{a2}}.

\begin{algorithm}
\caption{ECME Algorithm} \label{a2}
\begin{algorithmic}[1]
\STATE {Initialize $\mathbf{Q}^{(0)}\succ\mathbf{0}_N$, $K$, and $k=1$.}
        \WHILE{$k\le K$}
           \STATE {Obtain $\mathbf{S}^{(k)}[\mathbf{S}^{(k)}]^H$ using \eqref{15}.}
           \STATE {Obtain $\mathbf{Q}^{(k)}$ using \eqref{19}.}
           \STATE {$k=k+1$.}
       \ENDWHILE
\STATE {Let $\hat{\mathbf{S}}=\mathbf{S}^{(K)}$ and output $\hat{\mathbf{S}}$.}
\end{algorithmic}
\end{algorithm}

\subsection{Convergence and Complexity}

From \cite{liu,wu,McLachlan}, we know that since only the complex Gaussian distribution, whose PDF has a regular exponential-family form, is involved, the ECME algorithm satisfies the ``space-filling'' and other regularity conditions and its sequence $\big\{\big(\mathbf{S}^{(k)},\mathbf{Q}^{(k)}\big)\big\}$ always converges to a stationary point of $f(\mathbf{S},\mathbf{Q})$. Theoretically, $f(\mathbf{S},\mathbf{Q})$ has multiple stationary points, so an accurate $\mathbf{Q}^{(0)}$ should be required by the ECME algorithm. Fortunately, simulation results in \cite{stoica} and the subsequent section show that although the objective function $f(\mathbf{S},\mathbf{Q)}$ in problem \eqref{5} is non-convex, it seems to have a favourable landscape under moderately nonuniform noise, which makes its minimization by the FAAN method or the ECME algorithm a relatively straightforward task. However, since the second step at each iteration of the FAAN method sequentially (\textit{one by one}) updates the estimates of $\sigma_1,\dots,\sigma_N$ along the fixed coordinate directions, its sequence $\big\{\big(\mathbf{S}^{(k)},\mathbf{Q}^{(k)}\big)\big\}$ gets stuck at an undesired limit point more easily when the noise is immoderately nonuniform, which will be shown in \textbf{Fig. \ref{f11}}.

We utilize the following proposition to guarantee that the solution provided by the ECME algorithm is ``proper'' for problem \eqref{5} \cite{joreskog,adachi}.
\begin{proposition}
In the ECME algorithm, $\mathbf{S}^{(k)}[\mathbf{S}^{(k)}]^H\succeq\mathbf{0}_N$ and $\mathbf{Q}^{(k)}\succ\mathbf{0}_N$ for arbitrary $k\ge1$ if $\hat{\mathbf{R}}\succ\mathbf{0}_N$ and $\mathbf{Q}^{(0)}\succ\mathbf{0}_N$.
\end{proposition}

\begin{proof}
    We utilize the mathematical induction method. When $\hat{\mathbf{R}}\succ\mathbf{0}_N$ and $\mathbf{Q}^{(k-1)}\succ\mathbf{0}_N$, Eq. \eqref{15} directly indicates
    \begin{eqnarray}
        &\mathbf{S}^{(k)}[\mathbf{S}^{(k)}]^H\nonumber\\
        &=[\mathbf{Q}^{(k-1)}]^{1/2}\mathbf{U}^{(k)}\boldsymbol{\Lambda}^{(k)}[\mathbf{U}^{(k)}]^H[\mathbf{Q}^{(k-1)}]^{1/2}
        \succeq\mathbf{0}_N\nonumber
    \end{eqnarray}
    and then $\mathbf{C}^{(k)}=\mathbf{S}^{(k)}[\mathbf{S}^{(k)}]^H+\mathbf{Q}^{(k-1)}\succ\mathbf{0}_N$ and $\boldsymbol{\delta}^{(k)}=[\mathbf{C}^{(k)}]^{-1}\mathbf{Q}^{(k-1)}$ are of full rank. As a result, $[\boldsymbol{\delta}^{(k)}]^H\hat{\mathbf{R}}\boldsymbol{\delta}^{(k)}\succ\mathbf{0}_N\Rightarrow\hat{\mathbf{R}}^{(k)}_v=\boldsymbol{\Delta}^{(k)}+[\boldsymbol{\delta}^{(k)}]^H\hat{\mathbf{R}}\boldsymbol{\delta}^{(k)}\succ\mathbf{0}_N$ in \eqref{16e} due to the covariance matrix $\boldsymbol{\Delta}^{(k)}\succeq\mathbf{0}_N$. Finally, we obtain $\mathbf{Q}^{(k)}=\mathrm{diag}\big(\hat{\mathbf{R}}^{(k)}_v\big)\succ\mathbf{0}_N$ in \eqref{19}. The proof is completed.
\end{proof}

By comparing \textbf{Algorithm \ref{a1}} and \textbf{Algorithm \ref{a2}}, \textbf{Table \ref{t1}} and \textbf{Table \ref{t2}} give their main operations of the $k$th iteration. From both tables, we can know that the computational complexities of each iteration in both algorithms involve $\mathit{\mathcal{O}}(N^3)$ when $N$ is large \cite{Izadkhah}. Thus, the ECME algorithm has almost the same computational complexity at each iteration as the FAAN method.

\begin{table*}[h]
    \centering
    \caption{Computational complexity of the $k$th iteration in the FAAN method}
    \begin{tabular}{c | c | c | c}
    \hline
        \textbf{Main Operation} & \makecell{eigen-decomposition of\\$\tilde{\mathbf{R}}^{(k)}$ in \eqref{9}} & \makecell{$\boldsymbol{\Gamma}^{(k)}=\big(\mathbf{U}^{(k)}\boldsymbol{\Lambda}^{(k)}[\mathbf{U}^{(k)}]^H+\mathbf{I}_N\big)^{-1}$\\in \eqref{11}} & \makecell{the $b_n^{(k)}$'s\\in \eqref{11}} \\
    \hline
    \textbf{Complexity} &  $\mathit{\mathcal{O}}(N^3)$ &  $\mathit{\mathcal{O}}(N^3)$ &  $\mathit{\mathcal{O}}(K_2N^2)$ \\
    \hline
    \end{tabular}
    \label{t1}

    \centering
    \caption{Computational complexity of the $k$th iteration in the ECME algorithm}
    \begin{tabular}{c | c | c | c}
    \hline
       \textbf{Main Operation} & \makecell{eigen-decomposition of\\$\tilde{\mathbf{R}}^{(k)}$ in \eqref{15}} & $[\mathbf{C}^{(k)}]^{-1}$ in \eqref{16e} & \makecell{matrix multiplications in \eqref{16e}}\\
    \hline
    \textbf{Complexity} &  $\mathit{\mathcal{O}}(N^3)$ &  $\mathit{\mathcal{O}}(N^3)$ &  $\mathit{\mathcal{O}}(N^3)$ \\
    \hline
    \end{tabular}
    \label{t2}
\end{table*}

\section{Numerical Results}

Simulation results are used to show and compare \textbf{Algorithm \ref{a1}} and \textbf{Algorithm \ref{a2}}. Unless otherwise specified in the following figures, let $N=6$, $M=2$, $\theta_1=60^{\circ}$, $\theta_2=120^{\circ}$, $\mathbf{P}=10\times\mathbf{I}_M$, $\mathbf{Q}=\mathrm{diag}\{10,2,3,2,1,3\}$, $K_2=100$, $L=100$, $K=K_1=100$, and $\mathbf{Q}^{(0)}=\mathbf{I}_N$.

\subsection{Convergence Evaluation}

\begin{figure}[t] \centering
\includegraphics[scale=0.55]{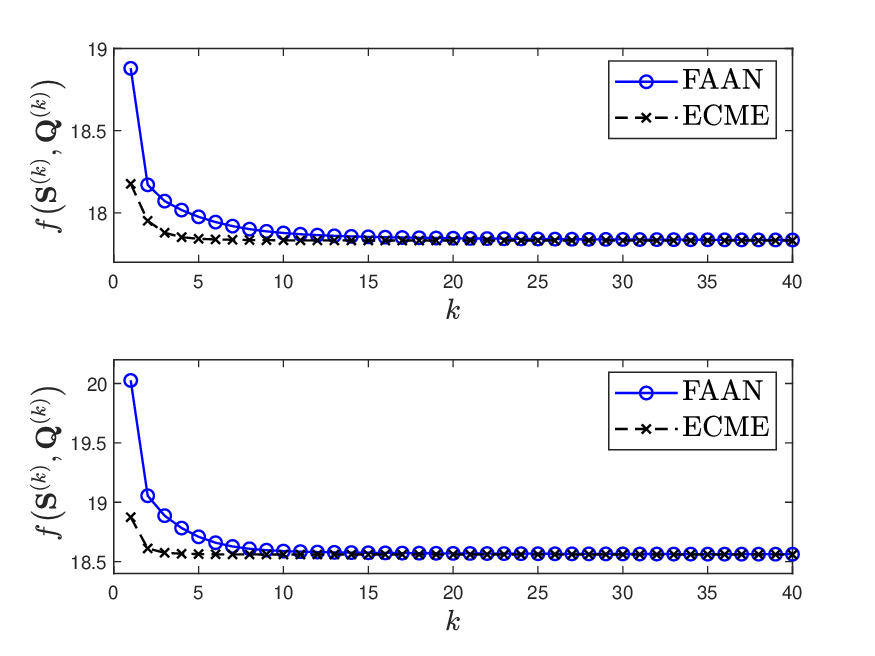}
\vspace{0cm}\caption{Convergence comparison of the FAAN method and the ECME algorithm\label{f1}}\vspace{0cm}
\end{figure}

Without loss of generality, \textbf{Fig. \ref{f1}} illustrates the convergence performances of both algorithms given the same snapshots. Here, the upper subfigure is with $\theta_1=90^{\circ}$ and $\theta_2=100^{\circ}$. We can clearly see that the ECME algorithm yields faster convergence than the FAAN method. Because the ECME algorithm has almost the same computational complexity at each iteration as the FAAN method, it is computationally more efficient.

After obtaining $\hat{\mathbf{S}}$ or the signal subspace estimate $\mathcal{C}(\hat{\mathbf{S}})$, we perform the eigen-decomposition of $\hat{\mathbf{S}}\hat{\mathbf{S}}^H$ to obtain $\hat{\mathbf{U}}_n\in\mathbb{C}^{N\times(N-M)}$, which consists of the eigenvectors corresponding to the $N-M$ smallest eigenvalues. To proceed, we adopt the orthogonal projector $\hat{\boldsymbol{\Pi}}_n=\hat{\mathbf{U}}_n\hat{\mathbf{U}}_n^H$\footnote{We do not adopt $\hat{\boldsymbol{\Pi}}_n=\mathbf{I}_N-\hat{\mathbf{S}}(\hat{\mathbf{S}}^H\hat{\mathbf{S}})^{-1}\hat{\mathbf{S}}^H$ since numerical results show its low accuracy.} onto the noise subspace estimate $\mathcal{C}(\hat{\mathbf{U}}_n)$ and the Root-MUSIC algorithm \cite{barabell} to obtain the DOA estimates $\hat{\theta}_m$'s.

\begin{figure}[t] \centering
\includegraphics[scale=0.55]{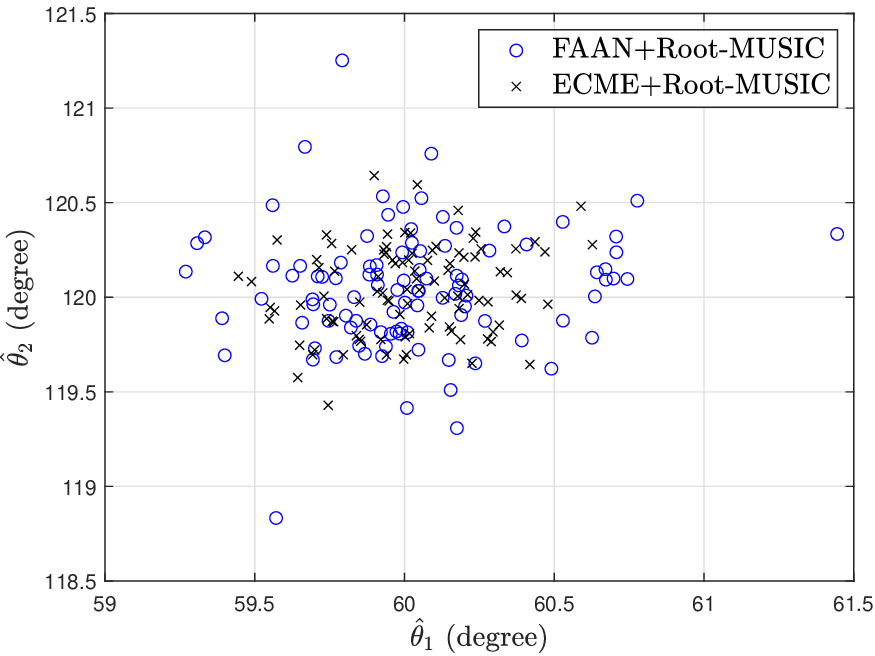}
\vspace{0cm}\caption{Scatter plot of $(\hat{\theta}_1,\hat{\theta}_2)$'s from both algorithms in moderately nonuniform noise\label{f2}}\vspace{0cm}
\end{figure}

\begin{figure}[t] \centering
\includegraphics[scale=0.55]{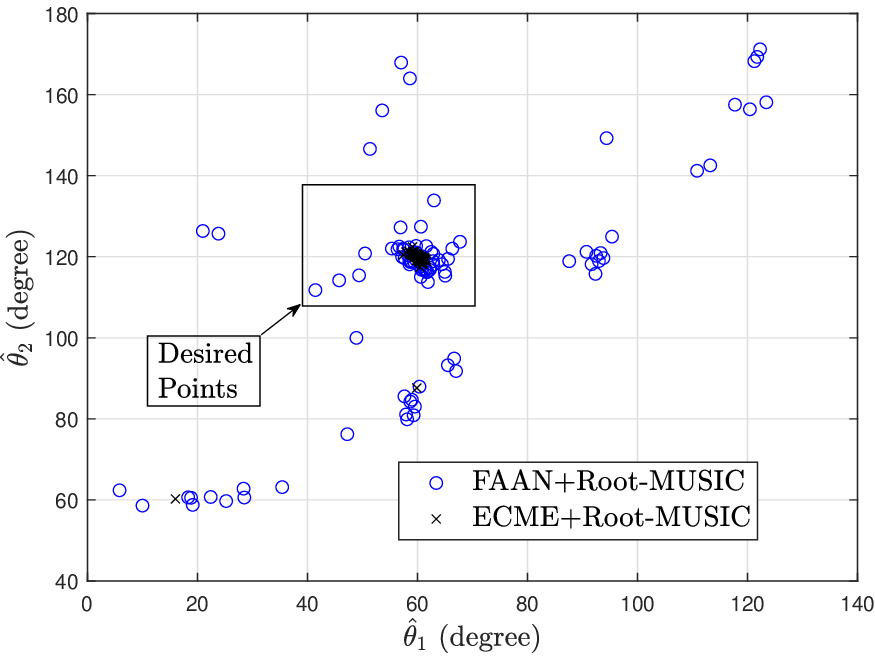}
\vspace{0cm}\caption{Scatter plot of $(\hat{\theta}_1,\hat{\theta}_2)$'s from both algorithms in immoderately nonuniform noise\label{f11}}\vspace{0cm}
\end{figure}

\textbf{Fig. \ref{f2}} and \textbf{Fig. \ref{f11}} show two scatter plots of $(\hat{\theta}_1,\hat{\theta}_2)$'s from both algorithms under $100$ independent realizations. The same snapshots of each realization are processed by both algorithms. In \textbf{Fig. \ref{f2}}, the noise is moderately nonuniform and all the $(\hat{\theta}_1,\hat{\theta}_2)$'s are centered on the true value $(60^{\circ},120^{\circ})$ since for each realization, both sequences $\{\mathbf{S}^{(k)}\mathbf{S}^{(k)}\}$'s from the FAAN method and the ECME algorithm converge to the solution of $\mathbf{S}\mathbf{S}^H$ in problem \eqref{5}. However, we know that under the same number of iterations $K=K_1$, both algorithms provide different $\hat{\mathbf{S}}\hat{\mathbf{S}}^H$'s for each realization, and the $\hat{\mathbf{S}}\hat{\mathbf{S}}^H$ provided by the ECME algorithm is closer to the solution of $\mathbf{S}\mathbf{S}^H$ in problem \eqref{5}. In \textbf{Fig. \ref{f11}}, $\mathbf{Q}=\mathrm{diag}\{10,2,3000,2,1,3\}$, the numbers of desired points from the FAAN method and the ECME algorithm are $53$ and $98$, respectively. This implies that when the noise is immoderately nonuniform, the convergence of the FAAN method to an undesired limit point is easier. That is to say, the convergence of the ECME algorithm to the global optimum is easier.

\begin{figure}[t] \centering
\includegraphics[scale=0.55]{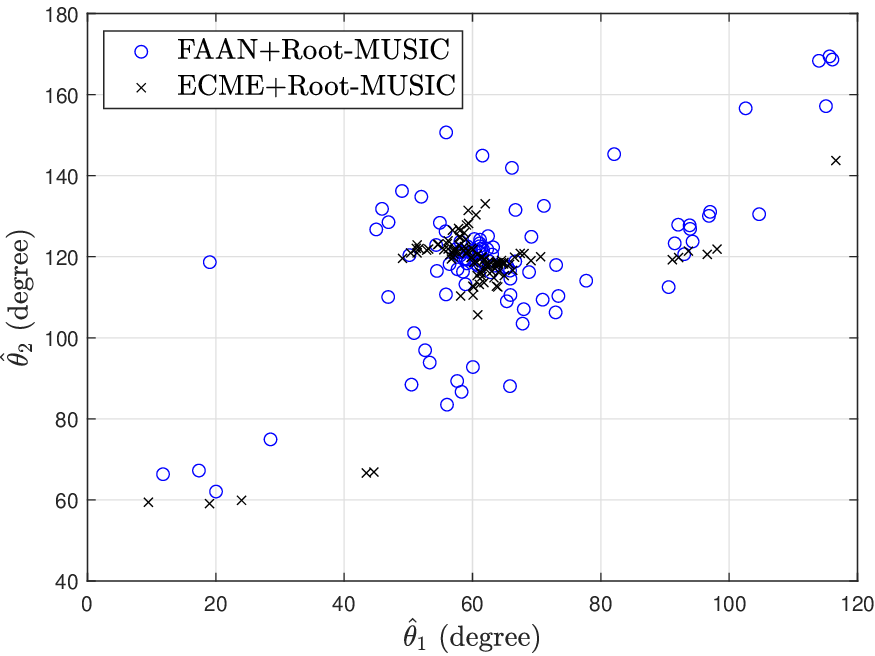}
\vspace{0cm}\caption{Scatter plot of $(\hat{\theta}_1,\hat{\theta}_2)$'s from both algorithms with coherent source signals\label{f4}}\vspace{0cm}
\end{figure}

When $\mathbf{P}=
\begin{bmatrix}
    10~10\\
    10~10
 \end{bmatrix}
$, i.e., the source signals are coherent, \textbf{Fig. \ref{f4}} shows a scatter plot of $(\hat{\theta}_1,\hat{\theta}_2)$'s from both algorithms under $100$ independent realizations. The same snapshots of each realization are processed by both algorithms. In this figure, many $(\hat{\theta}_1,\hat{\theta}_2)$'s from both algorithms are not centered on the true value $(60^{\circ},120^{\circ})$ although the noise is moderately nonuniform. This indicates that in the presence of coherent source signals, the classic MLFA problem \eqref{5} cannot be employed directly to estimate the signal subspace and the spatial smoothing technique should be considered \cite{Shan,Pillai}.

\subsection{Accuracy Evaluation}

In this subsection, the root mean square error (RMSE) performances of DOA estimation related to both algorithms are compared with the Cramer–Rao lower bound (CRLB) \cite{pesavento,gershman}. To guarantee the convergence of both algorithms to the global optimum, let the source signals be noncoherent and the noise be moderately nonuniform. In the following figures, each RMSE is based on $1000$ independent realizations and the RMSE of $\hat{\theta}_m$ is
\begin{eqnarray}
    \mathrm{RMSE}(\hat{\theta}_m)&=&\sqrt{\frac{1}{1000}\sum_{i=1}^{1000}\Big(\hat{\theta}_m^i-\theta_m\Big)^2}\nonumber\\
    &=&10\lg\Big[\frac{1}{1000}\sum_{i=1}^{1000}\Big(\hat{\theta}_m^i-\theta_m\Big)^2\Big]~\mathrm{(dB)},\nonumber
\end{eqnarray}
where $m=1,\dots,M$ and $\hat{\theta}_m^i$ denotes $\hat{\theta}_m$ obtained from the $i$th realization.

\begin{figure}[t] \centering
\includegraphics[scale=0.55]{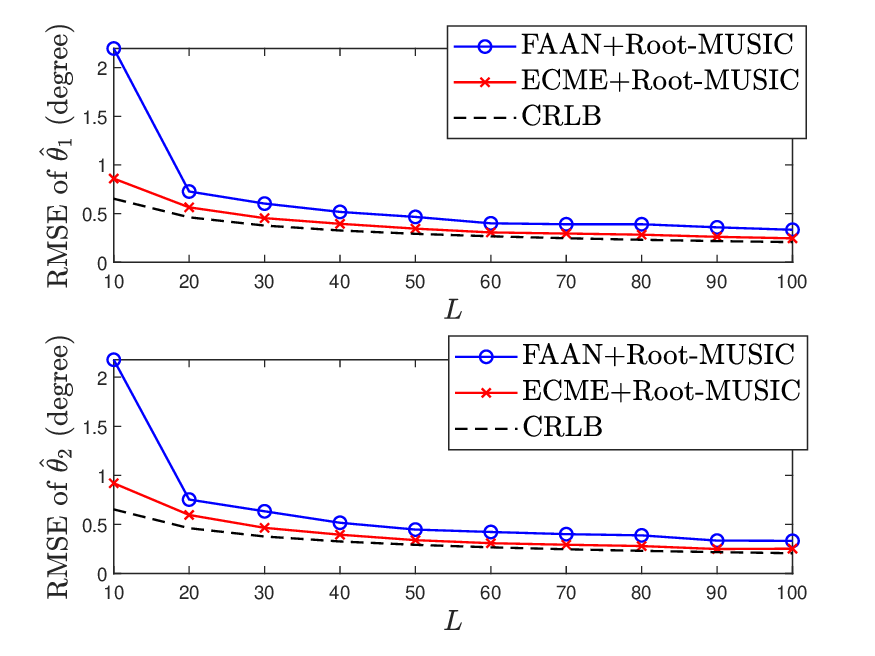}
\vspace{0cm}\caption{RMSE performances of $(\hat{\theta}_1,\hat{\theta}_2)$ from both algorithms versus $L$ \label{f3}}\vspace{0cm}
\end{figure}

\begin{figure}[t] \centering
\includegraphics[scale=0.55]{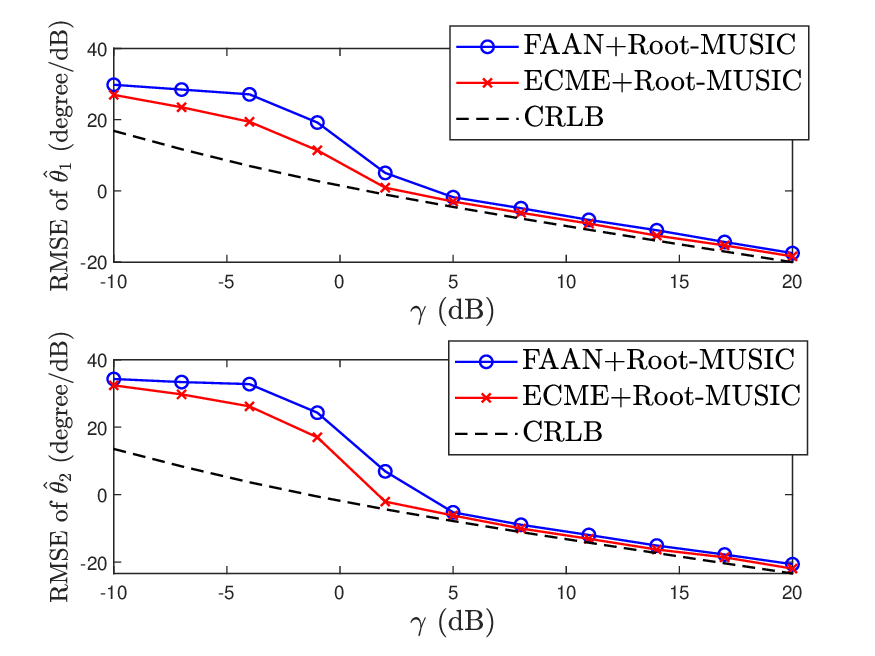}
\vspace{0cm}\caption{RMSE performances of $(\hat{\theta}_1,\hat{\theta}_2)$ from both algorithms versus $\gamma$ \label{f6}}\vspace{0cm}
\end{figure}

\begin{figure}[t] \centering
\includegraphics[scale=0.55]{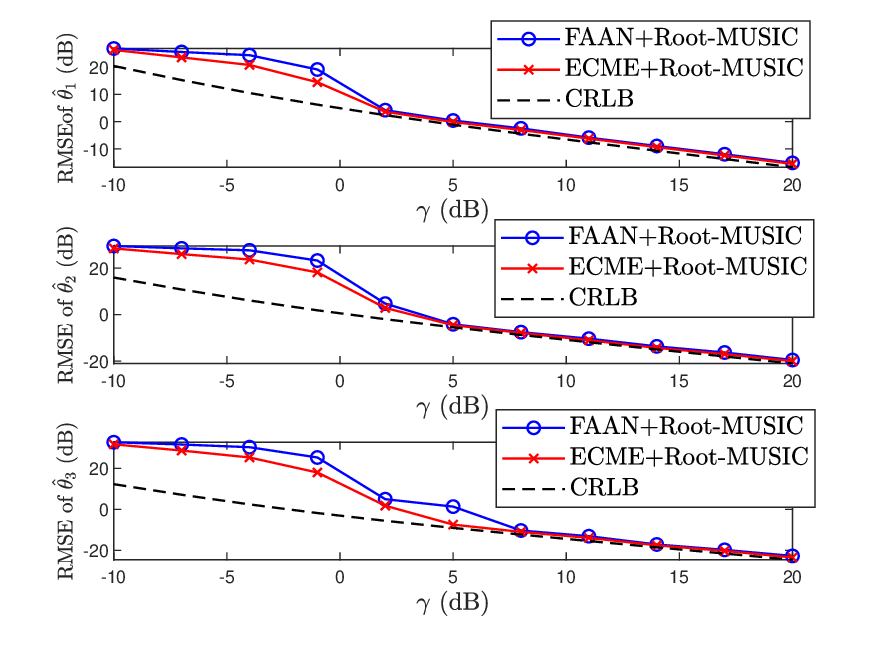}
\vspace{0cm}\caption{RMSE performances of $(\hat{\theta}_1,\hat{\theta}_2,\hat{\theta}_3)$ from both algorithms versus $\gamma$ \label{f7}}\vspace{0cm}
\end{figure}

\begin{figure}[t] \centering
\includegraphics[scale=0.55]{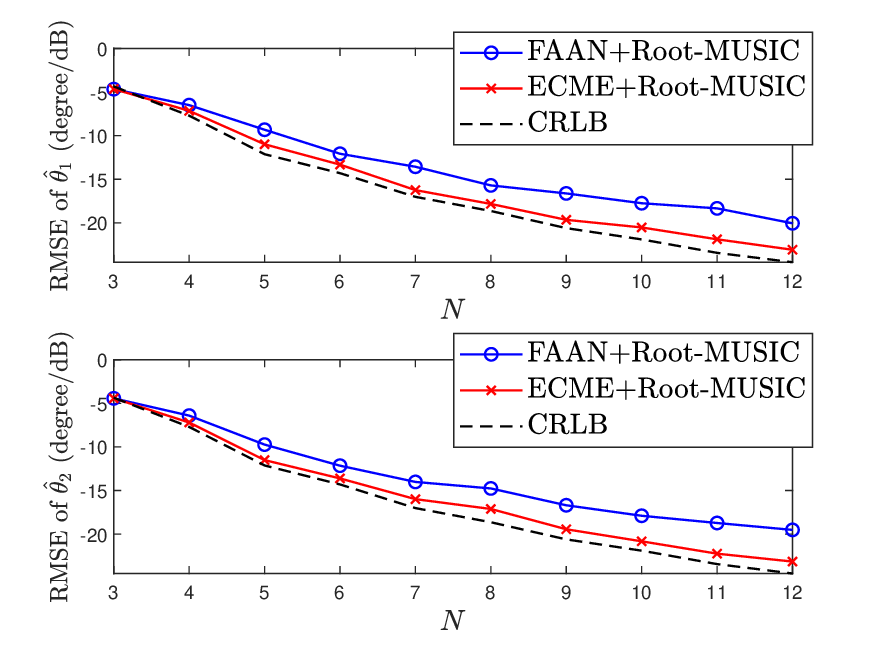}
\vspace{0cm}\caption{RMSE performances of $(\hat{\theta}_1,\hat{\theta}_2)$ from both algorithms versus $N$ \label{f8}}\vspace{0cm}
\end{figure}

\textbf{Figs. \ref{f3}}-\textbf{\ref{f8}} illustrate the RMSE performances of DOA estimation related to both algorithms versus $L$, $\gamma$, or $N$. In \textbf{Fig. \ref{f6}}, $\mathbf{P}=\gamma\times\mathbf{I}_M$, $\theta_1=40^{\circ}$, and $\theta_2=70^{\circ}$ while in \textbf{Fig. \ref{f7}}, $M=3$, $\theta_1=30^{\circ}$, $\theta_2=60^{\circ}$, $\theta_3=90^{\circ}$, and $\mathbf{P}=\gamma\times\mathbf{I}_M$. As expected, we observe that the RMSEs of DOA estimation from both algorithms approach the CRLBs as $L$ or $\gamma$ increases. More importantly, because the ECME algorithm yields faster convergence, its $\mathcal{C}(\mathcal{\hat{\mathbf{S}}})$ is more accurate under $K=K_1$ and its RMSE performance of DOA estimation is better. In \textbf{Fig. \ref{f8}}, $\sigma^2_n=6n/(N+1)$ for $(1/N)\sum_{n=1}^N\sigma^2_n=3$. This figure shows that the RMSEs of DOA estimation deviate from the CRLBs as $N$ increases, so MLFA cannot sufficiently capture DOA information from the snapshots $\mathbf{Y}$ when the number of sensors $N$ is large.

\begin{figure}[t] \centering
\includegraphics[scale=0.55]{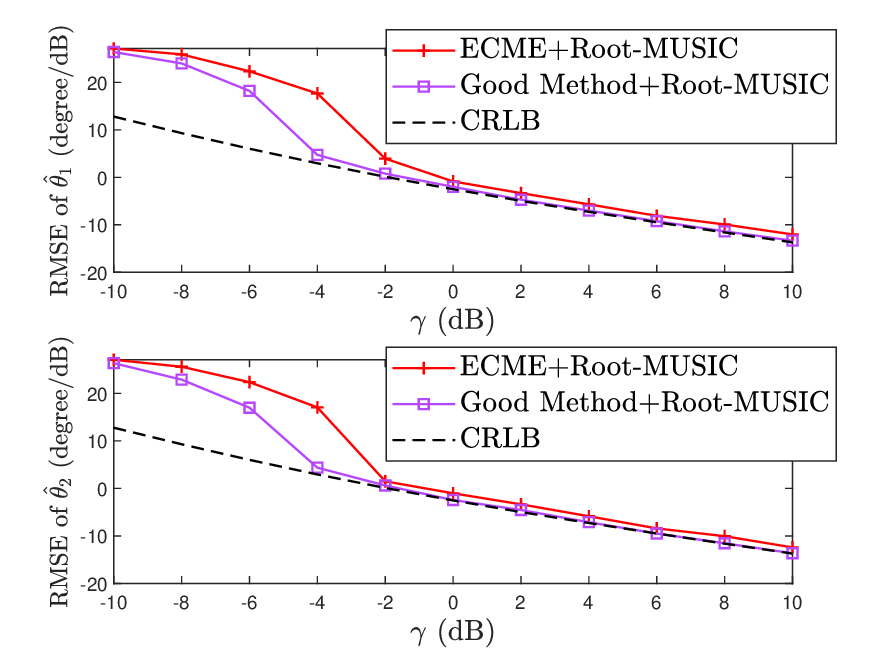}
\vspace{0cm}\caption{RMSE performances of $(\hat{\theta}_1,\hat{\theta}_2)$ from the ECME algorithm and a method versus $\gamma$ \label{f9}}\vspace{0cm}
\end{figure}

\begin{figure}[t] \centering
\includegraphics[scale=0.55]{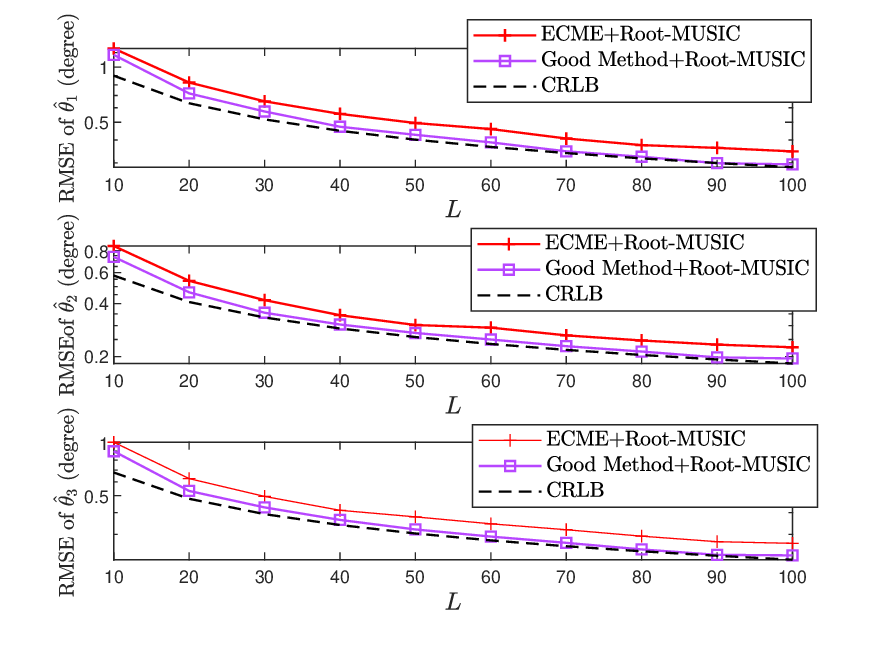}
\vspace{0cm}\caption{RMSE performances of $(\hat{\theta}_1,\hat{\theta}_2,\hat{\theta}_3)$ from the ECME algorithm and a method versus $L$ \label{f10}}\vspace{0cm}
\end{figure}

\textbf{Fig. \ref{f9}} and \textbf{Fig. \ref{f10}} compare the ECME algorithm with the state-of-the-art subspace separation approach in \cite{Majdoddin}, called ``Good Method''. In \textbf{Fig. \ref{f9}}, $\mathbf{P}=\gamma\times\mathbf{I}_M$ while in \textbf{Fig. \ref{f10}}, $M=3$, $\theta_1=40^{\circ}$, $\theta_2=80^{\circ}$, $\theta_3=120^{\circ}$. We can observe that each RMSE curve of DOA estimation from the ECME algorithm is above that from the ``Good Method'', so the noise subspace estimated by the ``Good Method'' is more accurate. Furthermore, because the ``Good Method'' only requires $5$ iterations and its computational complexity of each iteration also involves $\mathit{\mathcal{O}}(N^3)$, it is computationally more efficient. As a consequence, MLFA is not the best choice for the subspace based DOA estimation in unknown nonuniform noise\footnote{It is worth pointing out that the ECME algorithm is valuable since this algorithm should be the best choice for MLFA in other applications, e.g., psychometrics, econometrics, statistics and so on.}.

\section{Conclusion}

In this article, the ECME algorithm, an extension of the EM algorithm, has been designed again for MLFA by introducing new complete data, which can thus use two explicit formulas to sequentially update the estimates of parameters at each iteration and have excellent convergence properties. Theoretical analysis has shown that the ECME algorithm has almost the same computational complexity at each iteration as the FAAN method. However, numerical results have shown that the ECME algorithm yields faster stable convergence and the convergence to the global optimum is easier. Importantly, MLFA is not the best choice for the subspace based DOA estimation in unknown nonuniform noise.

\printbibliography 

\end{document}